\documentclass[aps, prl, superscriptaddress, reprint, nobibnotes]{revtex4-2}
\usepackage{amsmath,amssymb}
\usepackage{color}
\usepackage{comment}
\usepackage{bm,graphicx,hyperref}
\hypersetup{%
  breaklinks = {true},
  citecolor = {blue},
  colorlinks = {true},
  linkcolor = {red},}

\begin{document}

\title{Minimal model of drag in one-dimensional crystals}

\author{Harshitra Mahalingam}
\affiliation{Institute for Functional Intelligent Materials, National University of Singapore, 4 Science Drive 2, 117544, Singapore}

\author{Zhun Wai Yap}
\affiliation{Yale-NUS College, 16 College Avenue West, 138527, Singapore}
 
\author{B. A. Olsen}
\thanks{Corresponding author}
\affiliation{Yale-NUS College, 16 College Avenue West, 138527, Singapore}
\affiliation{Department of Physics, National University of Singapore, 2 Science Drive 3, 117551 Singapore}

\author{A. Rodin}
\thanks{Corresponding author}
\affiliation{Yale-NUS College, 16 College Avenue West, 138527, Singapore}
\affiliation{Centre for Advanced 2D Materials, National University of Singapore, 117546, Singapore}
\affiliation{Materials Science and Engineering, National University of Singapore, 117575, Singapore}

\begin{abstract}

Using a non-perturbative classical approach, we study the dynamics of a mobile particle interacting with an infinite one-dimensional (1D) chain of harmonic oscillators.
This minimal system is an effective model for many 1D transport phenomena, such as molecular motion in nanotubes and ionic conduction through solid-state materials. 
As expected, coupling between the mobile particle and the chain induces dissipation of the mobile particle's energy. 
However, both numerical and analytic results demonstrate an unconventional non-monotonic dependence of the drag on particle speed. 
In addition, when this system is subjected to a constant bias, it supports multiple steady-state drift velocities.

\end{abstract}	

\maketitle

\emph{Introduction.} In general, transport through a medium is accompanied by energy loss~\cite{Razavy}, leading to many familiar phenomena such as aerodynamic drag and friction.
Most classical formulations of dissipation, with the notable exception of friction, focus on liquid or gaseous media~\cite{Razavy, Chen2016}.
In these models, the size and energy scales of the moving object often significantly exceed those of the microscopic degrees of freedom (atoms and molecules of the medium).
These scales allow a series of simplifications, such as treating the medium as a continuous substance and regarding the drag force as Markovian~\citep{Zwanzig2001, Altland2010, Pathria2022}.

In contrast, motion through a solid occurs only when the size of the mover is comparable to the interatomic spacing of the medium. 
A relevant example is ionic transport through solids, which has been garnering attention recently in the context of solid-state batteries~\cite{Bachman2016, Manthiram2017, Famprikis2019} as a part of the search for more sustainable energy storage technologies~\citep{IEA2020, Yang2018}.
An important component of these batteries is the solid electrolyte: an electronic insulator that can conduct ions and acts as a separator between the anode and the cathode.

Dissipative motion through such crystalline solids precludes conventional approximations for two main reasons.
First, the mover's size (an atom or an ion in this case) can be smaller than the solid's interatomic spacing, making the continuum treatment of the medium inapplicable.
Second, long-range order in crystals, combined with similar time scales on which the mobile ion and the lattice atoms move, calls the Markovian approximation into question.
The most commonly used tools for studying ionic motion in solids are classical molecular dynamics (MD) and \textit{ab initio} molecular dynamics (AIMD) simulations~\citep{Wang2015, Deng2015, Kozinsky2016, DeKlerk2016, Adelstein2016, He2017, Krauskopf2017, Muy2018, Deng2017,  DiStefano2019}, though complementary approaches based on microscopic descriptions can be much less computationally intensive~\citep{Rodin2022, Rodin2022a}.

\begin{figure}
    \centering
    \includegraphics[width = \columnwidth]{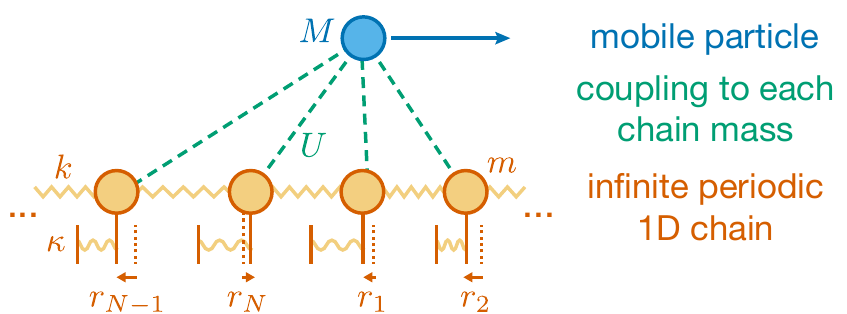}
    \caption{\emph{Schematic of the system.} A mobile particle of mass $M$ moves along a periodic 1D chain of $N\rightarrow \infty$ identical masses $m$, separated by distance $a$ at equilibrium.
    Each chain mass undergoes harmonic motion with spring constant $\kappa$ and displacement $r_g$, and couples to its neighbor with spring constant $k$.
    The $N$ oscillatory modes of the chain act as a bath, absorbing the mobile particle's energy due to the interaction $U$.}
    \label{fig:Schematic}
\end{figure}

The aim of these approaches is to develop intuition for the link between microscopic motion and macroscopic transport, in contrast to MD, which gives quantitative predictions of short-time phenomena.
The simplest realistic model of a crystal is a one-dimensional (1D) harmonic chain, as shown in Fig.~\ref{fig:Schematic}.
Our earlier work~\citep{Rodin2022a} demonstrated that such a chain can act as a heat bath for a particle confined to a harmonic trap.
This 1D crystal model is easy to study both analytically and numerically, and can also describe a range of experimentally relevant systems.
In some highly anisotropic ionic conductors, current-carrying ions travel along 1D channels~\citep{Furusawa1993, Furusawa2000, Mansson2014, Volgmann2017} or nanowires~\citep{Cho2022}, similar to motion in nanotubes of atoms~\citep{Lim2011, Senga2014, Tunuguntla2016, Hauser2017}, simple molecules~\citep{Khlobystov2005, Chen2019}, or fullerenes~\citep{Khlobystov2004, GimenezLopez2011}.

In this letter, we numerically demonstrate the emergence of drag in a 1D crystal. 
We also develop an analytic approach to relate the energy loss to the properties of the system in the absence of thermal motion.
In particular, we show that the interplay of geometry and the structure of the crystal lead to an unconventional dependence of drag on speed.
Specifically, the energy loss is non-monotonic in the velocity and decreases with higher speeds.
Furthermore, we predict that applying a bias to the system gives rise to multiple drift velocities, the values of which are determined by the crystal parameters.
Numerical confirmation of this prediction lends credence to our model as an appropriate tool for the problem, paving the way for subsequent studies focusing on higher dimensions and the role of thermal effects on transport.

We perform all our calculations using the {\scshape julia} programming language~\citep{Bezanson2017} and make our code available at https://github.com/rodin-physics/1d-chain-thermalization.
Due to their size, the output files are not included in the repository.
Our plots are visualized using Makie.jl,~\citep{Danisch2021} employing a color scheme suitable for color-blind readers, developed in \citep{Wong2011}.

\begin{figure}
    \centering
    \includegraphics{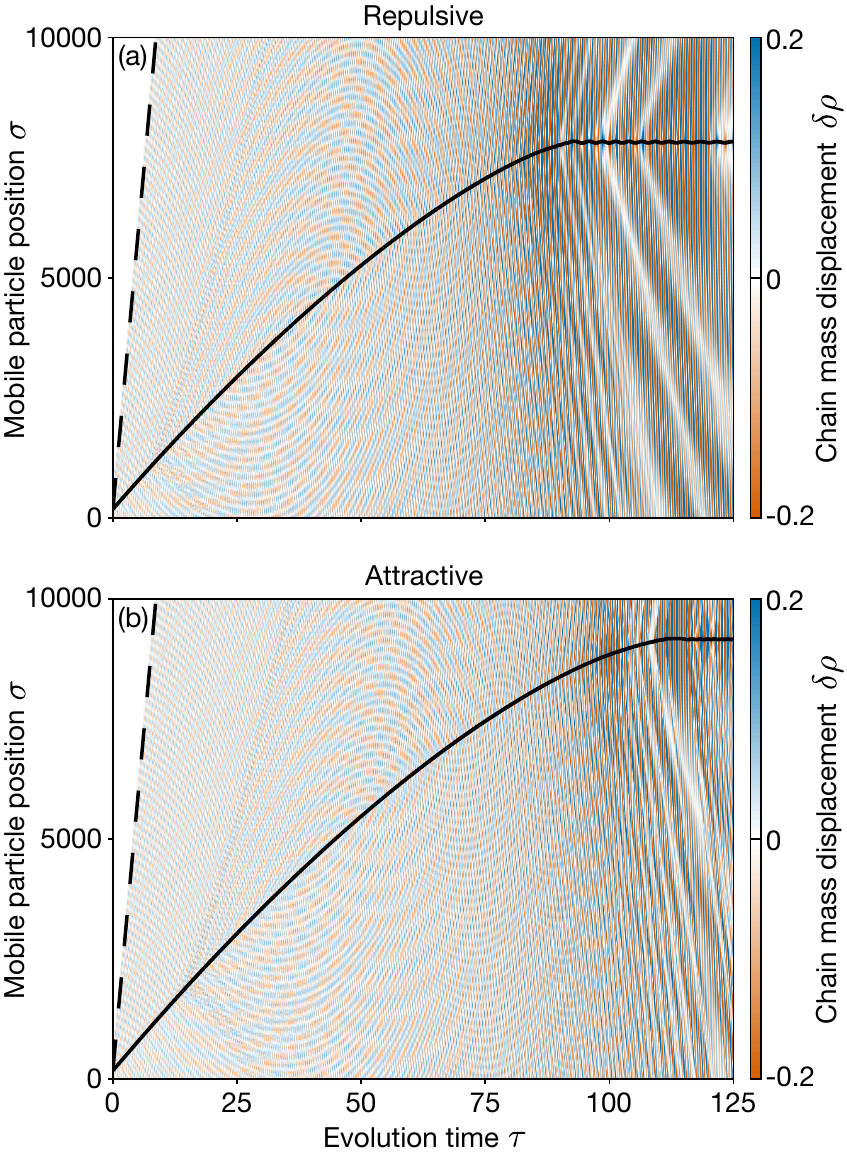}
    \caption{\emph{General example of dissipation.} 
    Motion of a single mobile particle with repulsive (a) and attractive (b) Gaussian chain-particle interaction.
    Both cases have chain spacing $\alpha = 40$, Gaussian width $\lambda = 4$ and amplitude $\Phi_0 = \pm 20$, $\mu = 1$, and initial mobile particle speed $\dot{\sigma}_0 = 120$ at a point halfway between chain masses.
    Both mobile particle trajectories $\sigma(\tau)$, shown in black lines, exhibit gradual slowing and eventual trapping in local energy minima. 
    The displacement of 250 individual chain masses $\delta\rho_j$ is shown with a heatmap.
    At late times, $|\delta\rho_j|$ reaches values as large as 0.8, so the colors are saturated.
    The dashed lines with slope $\pi\alpha(\omega_\mathrm{fast}-1)$ represent the edge of the sound cone, outside of which the chain masses are motionless.
    }
    \label{fig:General_Example}
\end{figure}

\emph{Model.} The 1D crystal in Fig.~\ref{fig:Schematic} hosts vibrational eigenmodes with frequencies

\begin{equation}
	\Omega_j 
	= \sqrt{\Omega_\mathrm{slow}^2+\left(\Omega_\mathrm{fast}^2-\Omega_\mathrm{slow}^2 \right)\sin^2\left(\frac{q_ja}{2}\right)}
	\label{eqn:Omega}
\end{equation}
and corresponding normalized eigenvectors $\boldsymbol{\varepsilon}_j$ with elements $\varepsilon_{g, j} = e^{i q_j a g} / \sqrt{N}$.
Here $q_j = 2\pi j / aN$ with $1\leq j\leq N$, where $1\leq g\leq N$ is the index of the chain mass.
The band edges $\Omega_\mathrm{fast} = \sqrt{4k/m+\kappa/m}$ and $\Omega_\mathrm{slow} = \sqrt{\kappa / m}$ are the maximum and minimum frequencies of the eigenmodes.

To guarantee that the system remains firmly in the classical regime, we will ensure that relevant length scales are larger than all the quantum lengths associated with the system.
The largest such quantity is the quantum oscillator length of the slowest oscillatory chain mode $l_\mathrm{slow} = \sqrt{\hbar / m\Omega_\mathrm{slow}}$.
Normalizing by $l_\mathrm{slow}$, we will express all lengths as dimensionless quantities.
Similarly, we express all frequencies in terms of $\Omega_\mathrm{slow}$, times in terms of $t_\mathrm{slow} = 2\pi / \Omega_\mathrm{slow}$, and energies in terms of $E_\mathrm{slow} = \hbar \Omega_\mathrm{slow}$.
We also express the mass of the mobile particle in terms of $m$ as $\mu = M / m$.

The time evolution of the system is governed by the following equations of motion (see SM Secs. I and II):

\begin{align}
\boldsymbol{\rho}(\tau)
   &= -2\pi \int^\tau d\tau'
   \tensor{\Gamma}(\tau - \tau')
   \nabla_{\boldsymbol{\rho}}\Phi\left[\boldsymbol{\rho}(\tau'),\sigma(\tau')\right]
   \,,
    \label{eqn:rho}
   \\
    \tensor{\Gamma}(\tau)
    &=\sum_j \boldsymbol{\varepsilon}_j \boldsymbol{\varepsilon}_j^\dagger\frac{\sin\left(2\pi\omega_j \tau\right)}{\omega_j}\,,
   \label{eqn:Gamma}
   \\
   \ddot{\sigma}(\tau) &= -(2\pi)^2\frac{1}{\mu}\sum_k\frac{d}{d\sigma_j}\Phi\left[\rho_k(\tau),\sigma(\tau)\right] \,.
   \label{eqn:sigma}
\end{align}
Here $\boldsymbol{\rho} = \mathbf{r} / l_\mathrm{slow}$ is the vector of all the (dimensionless) chain mass displacements, $\sigma$ is the position of the mobile particle, $\tau$ is the time, $\omega_j$ are the mode frequencies, and $\Phi$ is the pairwise interaction between the mobile particle and the chain masses.
The memory kernel $\tensor{\Gamma}(\tau)$ is a Toeplitz matrix whose $n$th diagonal is given by

\begin{align}
        \Gamma_{n}(\tau)
     &=
   \frac{2}{\pi}  \int_{0}^{\frac{\pi}{2}}d\theta \cos\left(2n\theta\right)
   \nonumber
   \\
   &\times\frac{\sin\left(2\pi \tau\sqrt{1+\left(\omega_\mathrm{fast}^2-1 \right) \sin^2\theta}\right)}{\sqrt{1+\left(\omega_\mathrm{fast}^2-1 \right)\sin^2\theta}}\,.
   \label{eqn:Gamma_Int}
\end{align}
Physically, Eq.~\eqref{eqn:Gamma_Int} acts as a propagator: following an impulse to the $l$th chain mass at time $\tau_0$, it describes the displacement of the $(l + n)$th chain mass at time $\tau_0 + \tau$ (see SM Sec. II for a visualization of this propagator).

\emph{Dissipation.} As a general example, we discuss a system with $\omega_\mathrm{fast} = 10$, interaction $\Phi$ with Gaussian shape of amplitude $\Phi_0=\pm 20$ and width $\lambda=4$, and lattice spacing $\alpha=40$.
We introduce a mobile particle with mass $\mu = 1$ and initial velocity $\dot \sigma_0 = 120$ midway between two chain masses.
As shown in Fig.~\ref{fig:General_Example}, the mobile particle dissipates kinetic energy, slowing until it eventually becomes trapped in a local energy well.
At this scale, it is hard to see, but the mobile particle's velocity undergoes fluctuations as it passes each chain mass.
For more details about velocity fluctuations, see SM Sec. III B.

We also display the displacement $\delta \rho_j$ of the chain masses near the mobile particle using a heatmap.
The amplitude of these displacements is very small compared to the chain mass spacing ($\delta \rho_j \ll \alpha$), and grows as the mobile particle slows.
Outside a sound cone the chain masses show no displacement.
This cone is well predicted by the greatest group velocity in the phonon band (see SM Sec. II for more detail).
After the mobile particle becomes trapped, the system undergoes persistent oscillation at frequencies just outside the phonon band, as discussed in Ref.~\cite{Rodin2022a}.

\emph{Dissipation Scaling.} Numerical integration of the equations of motion Eqs.~\eqref{eqn:rho}-\eqref{eqn:sigma} demonstrates dissipation, as expected.
To explore the dissipation's dependence on system parameters, we now develop an analytical model of energy loss when the mobile particle passes a single chain mass.
For an arbitrary interaction $\Phi$ which approaches zero at large separations, assuming the chain displacements vanish (justified by the small $\delta \rho$ seen in Fig.~\ref{fig:General_Example}), and taking the limit $\alpha \rightarrow \infty$, we obtain a single-pass dissipation
\begin{align}
   \Delta&=\sum_j
    \frac{1}{2N}\left[
    \frac{4\pi^2 \omega_j}{\dot{\sigma}^2}
     \int du\, \exp\left(i \frac{2\pi \omega_j}{\dot{\sigma}} u\right)\Phi(u)\right]^2
   \,.
   \label{eqn:general_loss_single_pass}
\end{align}
For a detailed derivation, see SM Sec. III A and C.
Based on the form of this equation, a stronger interaction between mobile particles and the chain results in a faster dissipation rate.

In the high-velocity limit, $\dot{\sigma} \rightarrow \infty$, the Fourier transform inside the square brackets approaches $\int dx\, \Phi(x)$ as the exponential function tends to 1.
If the integral is convergent, we find $\Delta \propto \dot{\sigma}^{-4}$.
Given that the frequency of encounters between the mobile particle and the chain masses is proportional to $\dot{\sigma}$, we multiply the energy loss per pass by the speed to obtain the energy dissipation rate: $\Delta \dot{\sigma} \sim \dot{\sigma}^{-3}$.
In between interactions with the chain masses, the mobile particle's total energy $\mathcal{E}$ is proportional to $\dot{\sigma}^2$, so the energy loss rate $\dot{\mathcal{E}}\sim - \mathcal{E}^{-3/2}$ produces a quasi-power law decay of the energy.
If the integral in Eq.~\eqref{eqn:general_loss_single_pass} is not convergent (such as in the case of Coulomb interaction), the divergence will be mitigated by the faster-decaying $\dot{\sigma}^{-4}$ prefactor, also leading to quasi-power-law energy decay.
Crucially, unlike typical drag, where faster motion produces more resistance by the medium, higher speed here actually results in reduced dissipation.

For small $\dot{\sigma}$, the behavior is strongly dependent on the potential profile.
For a non-diverging potential, the Fourier term in Eq.~\eqref{eqn:general_loss_single_pass} vanishes as small $\dot{\sigma}$ corresponds to a high Fourier momentum.
Consequently, for $\dot{\sigma} \lesssim \omega_j \Lambda$, where $\Lambda$ is the characteristic width of the potential, the chain mode does not absorb energy from the moving particle.
If $\dot{\sigma} \lesssim \omega_j \Lambda$ for all the modes in the chain, the particle experiences essentially no dissipation.
For singular potentials, the Fourier term diverges as a logarithm or a power of $2\pi\omega_j / \dot{\sigma}$, resulting in a {(quasi-)}power law dependence of $\Delta$ on $1 / \dot{\sigma}$, also leading to a power-law-like decay of energy with time.

\begin{figure}
    \centering
    \includegraphics[width = \columnwidth]{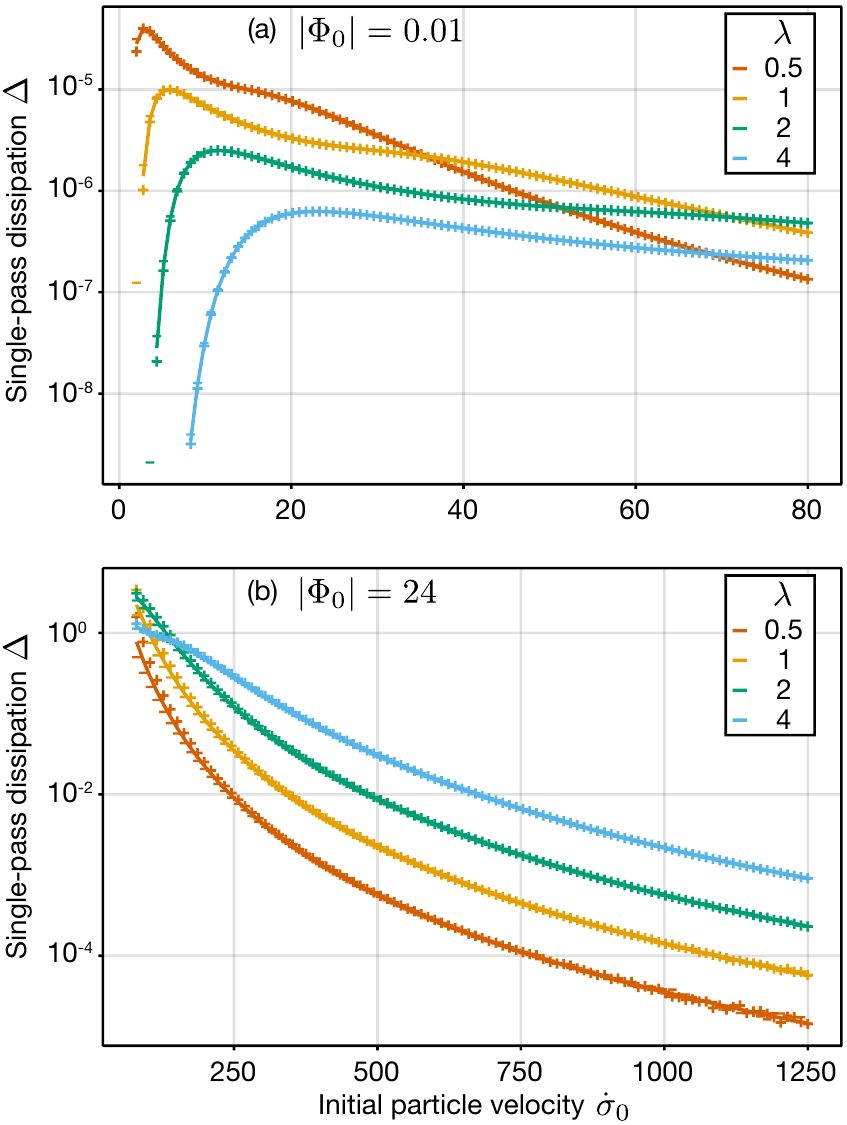}
    \caption{\emph{Single-pass dissipation.} Energy loss $\Delta$ for low velocity (a) with $|\Phi_0| = 1/100$, and high velocity (b) with $|\Phi_0| = 24$.
    The solid lines are analytical calculations using Eq.~\eqref{eqn:Delta} for different values of $\lambda$.
    The ``$+$''  and ``$-$'' symbols show the numerical values of $\Delta$ obtained from single-pass trajectories, where the symbols correspond to repulsive ($\Phi_0 > 0$) and attractive ($\Phi_0 < 0$) potentials respectively.
    }
    \label{fig:Dissipation}
\end{figure}

As a concrete example, we consider a Gaussian interaction, as in Fig.~\ref{fig:General_Example}, for which the sum in Eq.~\ref{eqn:general_loss_single_pass} can be computed analytically, yielding

\begin{align}
    \Delta &=
    4\pi^3\frac{\Phi_0^2}{\dot{\sigma}^2}
    \left(\frac{2\pi\lambda }{\dot{\sigma}}\right)^2
     e^{-\left(\frac{2\pi\lambda }{\dot{\sigma}}\right)^2 \left(\omega_\mathrm{fast}^2 +1 \right)/2}
     \nonumber
     \\
     &\times
     \left\{
     I_0(W)
     +\frac{\omega_\mathrm{fast}^2 -1}{2}\left[
     I_0(W)
     -
     I_1(W)
     \right]
     \right\}\,,
     \label{eqn:Delta}
\end{align}
where $W\equiv\left(\frac{2\pi\lambda }{\dot{\sigma}}\right)^2 \left(\omega_\mathrm{fast}^2 -1 \right)/2$,  and $I_n$ are modified Bessel functions of the first kind. 
In the fast mobile particle limit, $\dot{\sigma}\gg 2\pi\lambda \omega_\mathrm{fast}$,

\begin{equation}
    \Delta_\mathrm{fast} =
    2\pi^3\frac{\Phi_0^2}{\dot{\sigma}^2}
    \left(\frac{2\pi\lambda }{\dot{\sigma}}\right)^2\left(\omega_\mathrm{fast}^2 +1 \right)
    \,,
     \label{eqn:Delta_large_v}
\end{equation}
recovering the expected power law.
Conversely, for small values of $\dot{\sigma}$, the single-pass dissipation is exponentially suppressed:

\begin{equation}
    \Delta_\mathrm{slow} =
    4\pi^3\frac{\Phi_0^2}{\dot{\sigma}^2}
    \left(\frac{2\pi\lambda }{\dot{\sigma}}\right)
     e^{-\left(\frac{2\pi\lambda }{\dot{\sigma}}\right)^2 }
    \frac{1}{\sqrt{\pi \left(\omega_\mathrm{fast}^2 -1 \right)}}\,.
     \label{eqn:Delta_small_v}
\end{equation}
The analytic form of single-pass dissipation can be seen in Fig.~\ref{fig:Dissipation}, plotted for various interaction potentials in both low- and high-velocity limits.

\begin{figure*}[t]
    \centering
    \includegraphics[width = \textwidth]{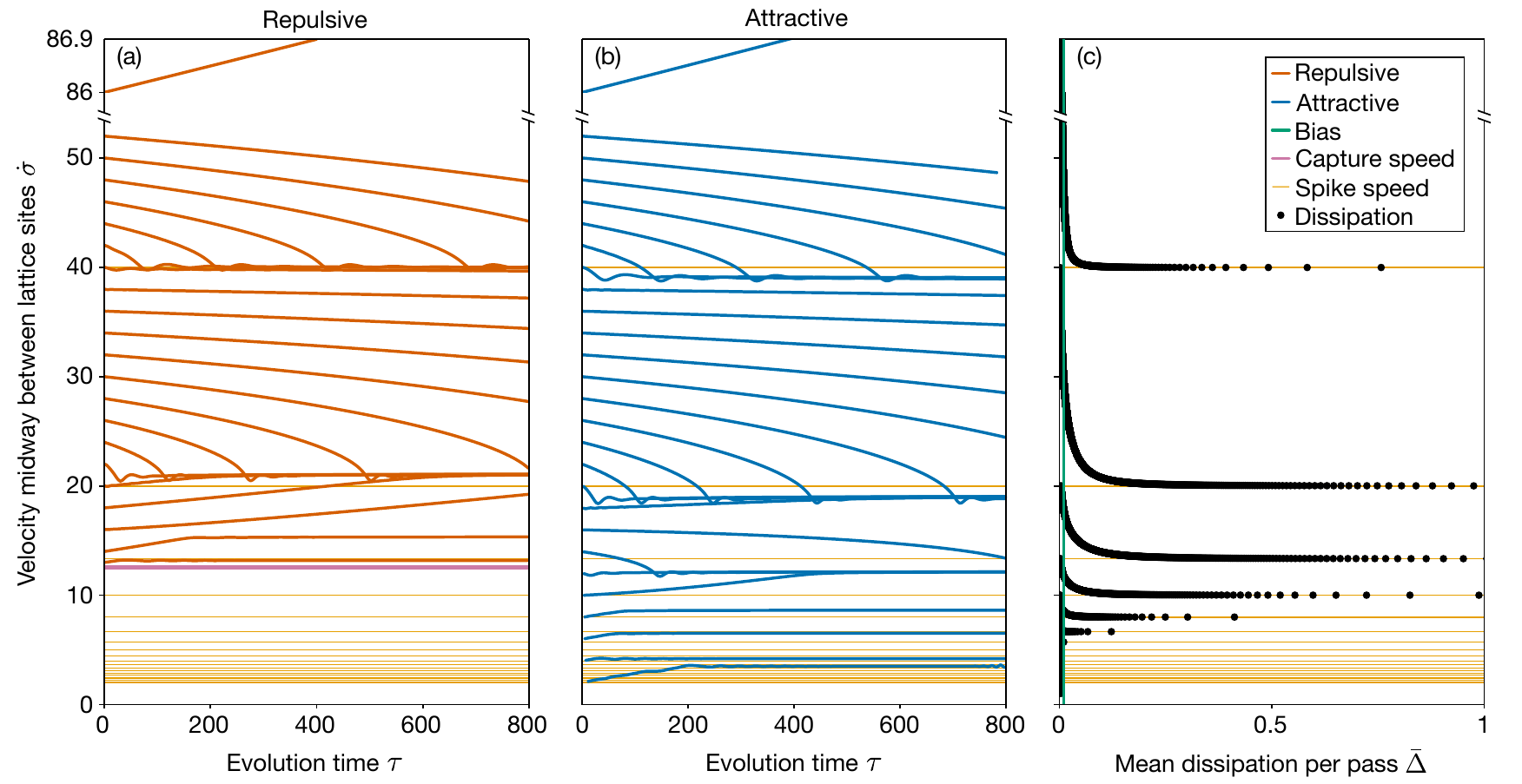}
    \caption{\emph{Multiple drift velocities maintained by constant bias.} The evolution of particle speeds for a range of initial speeds for system parameters $\alpha = 40$, $\Phi_0 = \pm 2$, $\lambda = 4$, and bias $\beta = 0.01$ for repulsive interactions (a) and attractive interactions (b). For clarity, we plot local maxima (minima) for repulsive (attractive) interactions and neglect the interaction-induced velocity fluctuation (see Sec. III B for details). The horizontal orange lines show velocities $\alpha / n$, where $n$ is a positive integer, corresponding to spikes in $\bar{\Delta}$. For the repulsive case, the minimum speed required to overcome the potential barrier (the capture speed) is given by $8 \pi^2 \Phi_0 / \mu \approx 12.6 $.
    Small bias leads to a very modest acceleration at high speeds, so we use a different scale to demonstrate a positive slope of the curve. (c) shows the mean dissipation per pass $\bar \Delta$ in the steady-state limit.}
    \label{fig:Drift}
\end{figure*}

To numerically validate the approximations made deriving Eq.~\eqref{eqn:general_loss_single_pass}, we initialize the mobile particle with various velocities $\dot \sigma_0$ halfway between two masses of a chain that is initially at rest.
We evolve the system for a time $\tau = 1.25 \times \alpha / \dot{\sigma}_0$ using the equations of motion \eqref{eqn:rho}--\eqref{eqn:sigma}.
This time is long enough for the mobile particle to pass a single chain mass, after which we calculate the kinetic energy $ T =M\dot{R}^2/2 = \hbar\Omega_\text{slow}(2\pi)^{-2} \mu\dot{\sigma}^2 /2$ and subtract it from the initial kinetic energy.
The resulting energy losses $\Delta$ agree well with the analytical predictions, as shown in Fig.~\ref{fig:Dissipation}.
For each set of parameters, the attractive ($\Phi_0<0$) and repulsive ($\Phi_0 > 0$) cases shift toward lower and higher $\dot\sigma_0$, respectively.
The analytic formula $\Delta(\dot{\sigma}_0)$ assumes a constant particle speed, which is not strictly true as the particle accelerates near the chain mass for $\Phi_0 < 0$ and decelerates for $\Phi_0 > 0$.
Consequently, the ``effective" speed that should enter in $\Delta(\dot{\sigma}_0)$ is greater than $\dot{\sigma}_0$ for the attractive case and smaller for the repulsive one.
This shift reflects the slight asymmetry of the two cases (see SM Secs. III B and III C).

We used different simulation parameters for the low- and high-$\dot{\sigma}$ regimes to avoid some technical issues.
For slow speeds, $\Phi_0$ must be smaller than the kinetic energy associated with the minimum value of $\dot{\sigma}$ to prevent the particle from getting stuck.
Using the same $\Phi_0$ for fast speeds is problematic, however, as $\Delta_\text{fast} \propto \Phi_0^2$ becomes an exceedingly small fraction of the initial kinetic energy, leading to numerical issues when subtracting $T(\tau)-T(0)$.
Hence, we calculated the low- and high-$\dot{\sigma}_0$ behavior using different values of $|\Phi_0|$.

\emph{Drift.} 
To study the long-time dynamics of the system, we calculate the energy dissipation per pass in the steady-state limit, when the mobile particle passes all $N$ chain masses, and $N\rightarrow\infty$ (see SM Sec. III E).
In this limit, only the chain modes with frequencies satisfying $\omega(x_s)= \sqrt{1 + (\omega_\mathrm{fast}^2 - 1)\sin^2\left(\pi x_s \right)} = -x_s\dot \sigma / \alpha$  absorb energy from the mobile particle, while the net energy exchange with all other modes vanishes. 
The mean dissipation per pass becomes
\begin{equation}
    \bar \Delta =
  \frac{1}{2}\sum_s
  \frac{\left[\frac{4\pi^2 \omega(x_s)}{\dot{\sigma}^2}
   \int du\, \exp\left(i \frac{2\pi \omega(x_s)}{\dot{\sigma}} u\right)\Phi(u)
   \right]^2}{\left|\alpha\omega'(x_s)/\dot{\sigma} + 1\right|} \, .
\end{equation}
When $-x_s \dot{\sigma} / \alpha \approx 1 = \omega_\text{slow}$, near the bottom band edge, the slowest phonon mode enhances the dissipation, leading to a spike in $\bar{\Delta}$, as seen in Fig.~\ref{fig:Drift}(c).
The speeds of these dissipation spikes are given by $\dot \sigma = \alpha / n$, for all positive integers $n$.

To emulate conduction in macroscopic devices, we additionally introduce a ``bias'' term in the form of a constant gradient in the potential experienced by the mobile particle.
This bias has the effect of increasing the mobile particle's kinetic energy by a fixed amount $\beta$ after each pass, which can counteract the energy $\bar \Delta$ dissipated to the chain. 
When $\bar \Delta(\dot \sigma) - \beta = 0$, a mobile particle will have velocity $\dot \sigma$ both before and after passing each chain mass. 
This behavior will lead to a constant average velocity, akin to a drift velocity in conducting materials.

As seen in Fig.~\ref{fig:Drift}(c), $\bar\Delta$ exhibits multiple spikes, suggesting the possibility of multiple bias-supported drift velocities. 
For a mobile particle moving with $\dot \sigma$ just above one of the spikes, a positive value of $\bar \Delta -\beta$ will cause the particle's velocity to decrease below the spike, and for velocity just below a spike, interaction with the chain and the bias will cause the particle's velocity to increase.
These behaviors lead to stable points in the vicinity of $\dot \sigma = \alpha / n$.
Corresponding to each spike is another, higher velocity that satisfies $\bar \Delta (\dot \sigma) - \beta  = 0$.
For velocities slightly higher (lower) than these solutions, the particle $\bar \Delta (\dot \sigma) - \beta $ is negative (positive), so the particle's velocity will increase (decrease), leading to an unstable repulsive point.

To confirm the unusual presence of multiple drift velocities, we perform numerical simulations of particle trajectories for a range of initial speeds, for both repulsive and attractive interactions in the presence of a bias.
Figure~\ref{fig:Drift}(a) and (b) plot the evolution of particle speeds with time and the spike locations, $\dot \sigma = \alpha / n$, are marked by horizontal lines.
We can see that the computed trajectories fall into three categories.
For high enough $\dot \sigma_0$, the dissipation is always smaller than the bias, and the particle accelerates away.
For intermediate velocities, the particle can accelerate or decelerate depending on the initial speed to a drift velocity near $\alpha/n$.
For the repulsive case, particles with $\dot \sigma_0$ below the capture velocity are trapped by the chain.
As predicted by our analytics, we see that multiple drift velocities are supported in both the repulsive and attractive cases. 

In reality, the particle does not move at a constant velocity throughout its trajectory, leading to moderate deviations from these predictions.
Near each chain mass, an attractive (repulsive) potential leads to an increase (decrease) in the particle velocity, so that the velocity takes a range of values (See Fig.~S2). 
This velocity fluctuation has the effect of broadening the spikes of Fig.~\ref{fig:Drift}(c) (See SM Sec. III E and SM Fig. S5).
This broadening will shift both the stable drift velocities and unstable solutions towards lower (higher) velocities for the attractive (repulsive) case.
In addition to shifting the spikes' locations, velocity fluctuations cause them to overlap, so that $\bar \Delta (\dot \sigma) - \beta = 0$ is not satisfied for low $\dot \sigma$.
For repulsive potentials, another drift velocity exists at low $\dot \sigma$ where the mobile particle almost comes to rest during each pass.
These solutions have velocity just above the capture velocity of the chain.
Additionally, at lower speeds, when the fluctuation magnitude becomes comparable to the drift velocity, the deviation from $\alpha/n$ is more pronounced.

Experimentally, an observed drift velocity in a 1D ionic conductor would serve as a probe of effective chain parameters. Since the highest $\dot \sigma_\text{drift}\approx\alpha= a / \sqrt{\hbar/\sqrt{\kappa m}}$, where the chain spacing $a$ can be directly measured, $\dot \sigma_\text{drift}$ indirectly measures the product of the chain mass and its confinement.
A realistic system, however, will experience thermal motion of the the chain masses, which will further blur the peaks in $\bar{\Delta}$. 
We will explore the role of these thermal fluctuations in future work.

\emph{Summary.}
We considered the 1D motion of a single mobile particle that interacts with each mass in an infinite chain via a nonlinear coupling.
We showed, using numerical simulations and a simplified single-pass model, that this interaction will dissipate the mobile particle's energy until it is trapped by the chain.
Unlike typical drag, this dissipation rate is reduced at higher speeds.
We also introduced a bias term, and using a steady-state model and numerical simulations, found that the resulting dynamics show a variety of behaviors.
The mobile particle can exhibit runaway acceleration, it can be trapped by the chain masses, but it can also settle into one of multiple stable drift velocities determined by chain parameters.
Signatures of these dynamics should be measurable in conduction channels where the motion is effectively 1D, such as electrons, simple molecules, and fullerenes in carbon nanotubes, or ions in anisotropic crystals.

\acknowledgments

A.R. acknowledges the National Research Foundation, Prime Minister Office, Singapore, under its Medium Sized Centre Programme and the support by Yale-NUS College (through Grant No. A-0003356-42-00).
B.A.O. acknowledges support from Yale-NUS College (through Grant Nos. A-0003356-39-00,  A-0000172-00-00, A-0000155-00-00, and C-607-261-026-001).
H.M. is supported by the Ministry of Education, Singapore, under its Research Centre of Excellence award to the Institute for Functional Intelligent Materials (I-FIM, project No. EDUNC-33-18-279-V12).
The computational work involved in this project was partially supported by NUS IT Research Computing Group.

B.A.O. and A.R. conceptualized the work; H.M. and A.R. wrote the code which was used by them and Z.W.Y. to run the simulations; H.M., B.A.O., and A.R. analyzed the results, prepared the graphics, and wrote the manuscript.


%

\end{document}